\documentclass[a4paper]{spie}  

\usepackage{amsmath,amsfonts,amssymb}
\usepackage{graphicx}
\usepackage[colorlinks=true, allcolors=blue]{hyperref}
\usepackage{astro_bib_macro}

\usepackage{booktabs}

\usepackage{listings}

\usepackage{enumitem}
\newlist{arrowlist}{itemize}{1}
\setlist[arrowlist]{label=$\Rightarrow$}

\title{First high-contrast results on THD2 testbed after infrastructure upgrade}

\author{Iva Laginja\supit{a,b,c}, Pierre Baudoz\supit{a}, Rémi Soummer\supit{d}, Emiel H. Por, Axel Potier\supit{a}, Johan Mazoyer\supit{a}, Raphaël Galicher\supit{a}, David Doelman\supit{b,f}, Rico Landman\supit{b}, Arnaud Sevin\supit{a}, Erin Pougheon\supit{a}, Corentin Paviot\supit{a}, Frans Snik\supit{b}, Felix Bettonvil\supit{b}, Jeroen Rietjens\supit{f}, Chris van Dijk\supit{g}, Kristien Peeters\supit{g}, Alexander Eigenraam\supit{g}, Mariya Krasteva\supit{h}, Matteo Taccola\supit{i}
\skiplinehalf
\supit{a} LIRA, Observatoire de Paris, Université PSL, CNRS, Sorbonne Université, Université Paris Cité, 5 place Jules Janssen, 92195 Meudon, France\\
\supit{b} NOVA/Leiden University, Einsteinweg 55, 2333 CC Leiden, The Netherlands\\
\supit{c} Universit\'e C\^ote d'Azur, Observatoire de la C\^ote d'Azur, CNRS, Laboratoire Lagrange, Nice, France\\
\supit{d} Space Telescope Science Institute, 3700 San Martin Drive, Baltimore, MD 21218, USA\\
\supit{e} Department of Astronomy and Astrophysics, University of California, Santa Cruz, CA, USA\\
\supit{f} SRON, Space Research Organisation Netherlands, Niels Bohrweg 4, 2333 CA Leiden, The Netherlands\\
\supit{g} cosine Remote Sensing B.V., Warmonderweg 14, 2171 AH Sassenheim, The Netherlands\\
\supit{h} Aix Marseille Université, CNRS, CNES, LAM, Marseille, France\\
\supit{i} European Space Agency, ESTEC, Keplerlaan 1, 2200 AG Noordwijk, the Netherlands\\
}

\authorinfo{Further author information, send correspondence to Iva Laginja: E-mail: iva.laginja@oca.eu}

\begin{document} 
\maketitle

\begin{abstract}
We present the first scientific results delivered by the upgraded THD2 high-contrast imaging testbed. We report two advances enabled by its improved stability and broadband performance. First, for the Roman Space Telescope, we demonstrate that Gaussian-shaped diversity probes outperform the baseline sinc probes by reducing non-linearities, supporting higher probe amplitudes, and improving electric-field estimation efficiency. These results have led to their prioritization as an enhanced early observation for Roman. Second, within ESA’s SUPPPPRESS project, we test new polarization-independent Vector Vortex Coronagraphs and design them to high-contrast performance approaching $10^{-10}$ over a $>20$\% bandwidth. We assess their behavior in narrow- and broad-band light with active focal-plane wavefront control. Together, these results show how THD2 strengthens Europe’s capability in high-contrast imaging, providing a unique platform reaching contrasts of $10^{-8} - 10^{-9}$  for developing next-generation coronagraphic technologies.
\end{abstract}


\section{Introduction}
\label{sec:introduction}

\subsection{Scientific Context: Direct Imaging of Exoplanets}
\label{subsec:scientific-context}

Direct imaging provides access to the photons emitted or reflected by an exoplanet by spatially separating planet from star light. This makes it possible to obtain spectra and constrain the planets' atmospheric composition, cloud properties, temperature, surface gravity, and chemical disequilibrium. These measurements inform theories of planet formation and evolution and, for temperate terrestrial planets, may eventually enable searches for atmospheric signatures associated with habitability\cite{Schwieterman2024AnOverview}.

The central observational difficulty is the combination of a very large star-to-planet flux ratio, or astrophysical contrast, and a very small angular separation. High-contrast coronagraphic instruments address this problem through a sequence of complementary techniques. A coronagraph suppresses the deterministic diffraction pattern of the on-axis star, while one or more deformable mirrors (DMs) are used in closed loop to estimate and correct residual aberrations, and to create a region of reduced stellar intensity, referred to as a dark hole (DH). Residual quasi-static and temporally varying speckles are then further attenuated or distinguished from astrophysical signals through post-processing. Although post-processing is essential to the final detection sensitivity of an astronomical instrument, the present work focuses on the coronagraphic and active wavefront-control layers.

A representative present-day observational regime is a contrast of approximately $10^{-5}$--$10^{-6}$ at angular separations near one arcsecond, whereas the direct characterization of Earth analogs around nearby Sun-like stars requires performance approaching $10^{-10}$ at separations of a few tens of milliarcseconds. Bridging this gap requires progress in coronagraph design, wavefront sensing and control, optical stability, calibration, polarization control, and broadband operation. The Extremely Large Telescope (ELT) on the ground, with its first generation of coronagraphic instruments, and the Nancy Grace Roman Space Telescope (Roman) from space provide important intermediate steps toward this regime. In particular, the Roman Coronagraph Instrument is intended to demonstrate active coronagraphic wavefront control in space at contrasts of order $10^{-7}$\cite{Mennesson2022TheRomanSpace}, while the Habitable Worlds Observatory\cite{Feinberg2026HabitableWorlds} (HWO) motivates the longer-term development of systems capable of substantially deeper contrast at smaller angular separations.

\subsection{The Role of High-Contrast Imaging Testbeds}
\label{subsec:role-of-hci-test-beds}

The technologies required to reach these contrast levels must be validated at the system level before they can be deployed on large ground-based or space-based observatories. Component-level characterization alone cannot fully capture the interaction between coronagraph masks, DMs, optical propagation, detector response, polarization effects, as well as the impact of model errors, temporal drifts, and control algorithms. High-contrast imaging (HCI) testbeds therefore provide controlled environments in which complete wavefront-sensing and control (WFS\&C) loops can be exercised repeatedly, their limitations can be isolated, and new observing concepts can be compared under reproducible conditions.

Different facilities address complementary aspects of this problem through distinct optical layouts, environmental conditions, coronagraph inventories, and mission priorities. Within this landscape, the THD2 testbed\cite{Baudoz2024Polarization,Baudoz2018OptimizationPerformanceMultideformable} at LIRA/Paris Observatory-PSL is a mature in-air platform for the development and validation of HCI technologies. Its long operational history makes it particularly valuable for comparative experiments: the principal sources of instability, polarization effects, and parasitic reflections are well characterized, while the optical configuration remains sufficiently flexible to accommodate new components and control strategies.

\subsection{The THD2 Testbed}
\label{subsec:thd2-test-bed}

THD2 has been developed and operated since 2015, based on the original work of THD that started in 2010. The testbed includes two DMs with $34\times34$ and $32\times32$ actuators, a tip--tilt mirror, and a tunable broadband source. It supports several focal-plane WFS\&C approaches, including pair-wise probing with electric-field conjugation\cite{Giveon2007ClosedLoopDM} (EFC), the self-coherent camera\cite{Baudoz2005TheSelfCoherentCamera} (SCC), and implicit EFC\cite{Haffert2023ImplicitElectric}, and it has hosted a broad range of coronagraphs, including four-quadrant phase masks (FQPMs\cite{Bonafous2016DevelopmentAndCharacterization}), hybrid Lyot coronagraphs\cite{Trauger2012ComplexApodizationLyot,Kuchner2002ACoronagraphWithABandLimited} (HLCs), and vector vortex coronagraphs\cite{Mawet2009OpticalVectorial} (VVCs). Previous work demonstrated monochromatic in-air contrasts below $10^{-8}$\cite{Potier2020_THD2} and established THD2 as a platform for both coronagraph development and control-algorithm validation. More detailed descriptions of the testbed, its historical performance, and the Roman-like configurations used in later studies are available in Refs. Laginja et al. (2025)\cite{Laginja2025ExtendedLinearity}, Baudoz et al. (2024)\cite{Baudoz2024Polarization} and Baudoz et al. (2018)\cite{Baudoz2018OptimizationPerformanceMultideformable}.

Despite this strong optical heritage, the original control infrastructure had become increasingly difficult to maintain. Several devices were distributed across aging computers, parts of the control stack depended on commercial or outdated software, and the resulting architecture limited  reproducibility, portability, and access by new users. The central goal of the upgrade described in this paper was therefore not simply to replace obsolete hardware, but to preserve the demonstrated optical performance of THD2 while establishing a modern, open-source, version-controlled control environment suitable for future experiments and international collaborations.

The remainder of this paper describes the migration of THD2 to the CATKit2 control framework\cite{Catkit2_2026} and its validation with the baseline FQPM testbed configuration. We then present two examples of science enabled by the upgraded infrastructure. The first places previously published by Laginja et al. (2025)\cite{Laginja2025ExtendedLinearity} Roman-like HLC and alternate-probing experiments in the context of ongoing work within the Roman Coronagraph Community Participation Program, including more recent developments based on
Gaussian probes. The second summarizes broadband tests of liquid-crystal VVCs performed within the ESA-funded SUPPPPRESS project, published by Landman \& Doelman et al. (2026)\cite{LandmanDoelman2026SUPPPPRESS}. These case studies illustrate how the upgraded infrastructure has expanded the range, efficiency, and collaborative scope of experiments that can be carried out on THD2.

\section{Software and Control Infrastructure Upgrade}
\label{sec:software-control-upgrade}

\subsection{Design Philosophy for Modern HCI Testbeds}
\label{subsec:design-philosophy}

A HCI testbed ideally remains useful beyond the lifetime of any individual experiment. Its control infrastructure must therefore support not only a predefined sequence of measurements, but also the integration of new hardware, coronagraphs, calibration procedures, and wavefront-control algorithms. This requirement is particularly important for mature facilities such as THD2, where the scientific value of the optical system extends across successive projects and collaborations. A control system optimized around one hardware configuration or one experimental sequence can produce successful results in the short term, but it becomes increasingly difficult to maintain when devices are replaced, software dependencies become obsolete, or new users need to reproduce earlier experiments or design new ones.

Modern testbed operation consequently benefits from a separation between device-specific control and higher-level experimental logic. Such a structure makes it possible to modify individual hardware components without rewriting complete experiments and provides a stable interface through which different projects can share acquisition, calibration, and diagnostic tools. More generally, encoding routine procedures in reusable software reduces dependence on project-specific scripts and preserves operational knowledge as personnel and scientific priorities evolve.

\subsection{The CATKit2 Hardware Control Library}
\label{subsec:catkit2-hardware-control}

CATKit2 is an open-source hardware-control framework developed from the operational experience of the HiCAT testbed \cite{Catkit2_2026}. It combines low-level services implemented in C++ with a Python-facing interface used to configure and execute experiments. Communication between operational units, so-called services, is based on shared-memory data streams, allowing hardware processes, monitoring tools, and experiment scripts to exchange commands and data while remaining functionally independent. This architecture retains the performance and direct hardware access of compiled services while providing an interface suited to the Python-based analysis and control tools commonly used by the astronomical community.

Each hardware component is controlled through a dedicated service that exposes a standardized set of commands and data products. Higher-level scripts therefore interact with cameras, DMs, light sources, and positioning stages through a common control layer rather than through vendor-specific software. The hardware services can continue operating independently of a particular experiment script, which facilitates monitoring, error handling, and recovery from interrupted measurements. The same infrastructure also supports graphical interfaces for testbed status, hardware operation, calibration, and closed-loop experiments.

\begin{figure}
    \centering
   \includegraphics[width = \textwidth]{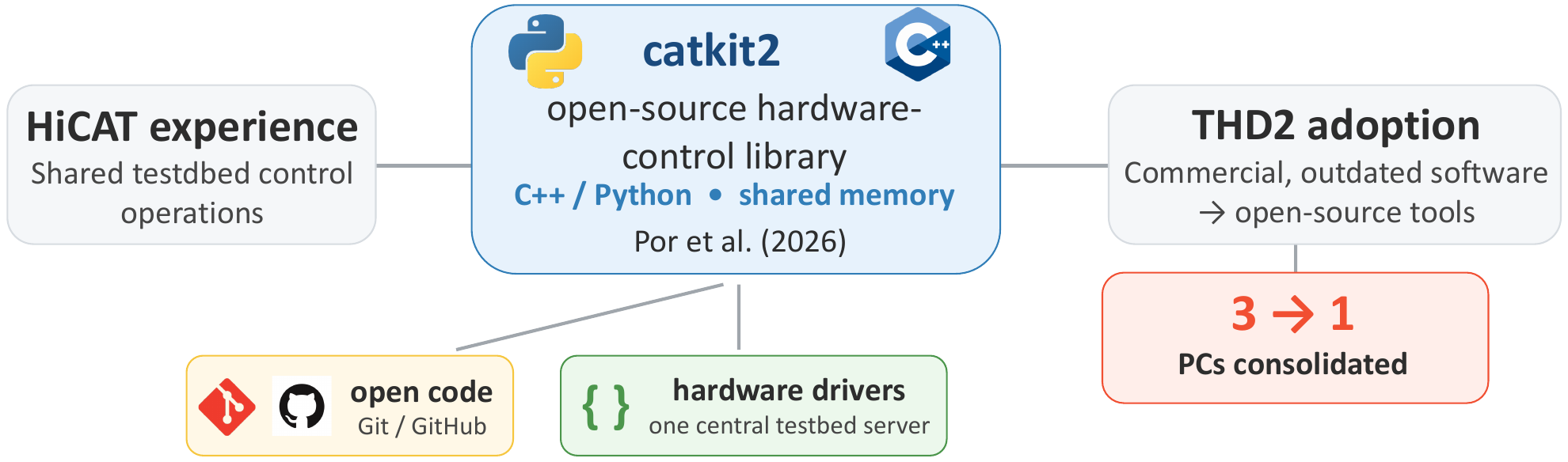}
   \caption[Figure] 
   {\label{fig:catkit2_upgrade} 
   Adoption of the CATKit2 hardware-control framework on THD2. Experience gained through shared testbed operations on HiCAT informed the open-source CATKit2 architecture, which combines C++ hardware services, a Python application programming interface, and shared-memory communication. On THD2, the migration replaced commercial, outdated software with version-controlled code and device-specific hardware services, while consolidating control processes previously distributed across three computers onto a single central server.}
   \end{figure}

The framework is developed using Git-based version control and includes openly accessible source code. This makes modifications traceable and allows improvements made for one testbed to be reviewed, reused, or adapted by other facilities. Figure~\ref{fig:catkit2_upgrade} summarizes the transfer of this architecture from the HiCAT development context to THD2.
These characteristics make CATKit2 particularly well suited to the modernization of THD2. Its separation of hardware services from experiment logic allows the legacy device interfaces to be replaced incrementally while preserving the higher-level structure of calibration and wavefront-control experiments. Its Python interface provides a common environment for experiment development, and supports reproducibility and collaborative maintenance. For THD2, the adoption of CATKit2 replaces commercial, outdated control software and consolidates processes that had previously been distributed across three computers onto a single one.

\subsection{Adapting CATKit2 to THD2}
\label{subsec:adapting-catkit2}

The upgraded THD2 software follows the three-layer architecture described by Soummer et al. (2026)\cite{Soummer2026}. CATKit2 provides the generic hardware-control layer, including device communication, shared-memory streams, synchronization, and experiment orchestration. CATKit2-HCI provides reusable HCI functionality, such as pair-wise probing, electric-field estimation, EFC, calibration routines, diagnostics, and standardized outputs. The private \texttt{thd-controls} package contains the THD2-specific layer: hardware configurations, bench conventions, calibration products, operating procedures, and experiment scripts. In short, CATKit2 controls the devices, CATKit2-HCI supplies reusable HCI methods, and \texttt{thd-controls} defines how both are used on THD2.

Adapting this architecture required adaptation of services for the two DMs (both by Boston Michromachines) and tunable light source (SuperK Fianium, NKT Photonics), and dedicated services for the science (Hamamatsu ORCA Quest 2) and auxiliary cameras (Allied Vision), and motorized stages. Each device is represented by an independent CATKit2 service that exposes commands, status information, and data streams through a common interface. Python experiment scripts therefore operate the testbed through CATKit2 rather than through vendor-specific software. Because services remain active independently of an experiment script, hardware states and image streams can also be monitored continuously and recovered more easily after an interrupted measurement.

The corresponding THD2 service definitions are maintained in \texttt{thd-controls}. During the migration, device identifiers, communication parameters, reference positions, command limits, initialization sequences, and safety constraints were translated into CATKit2 configurations. Bench-specific quantities, including DM geometries, camera orientations, DH definitions, and calibration products, also remain in this package. Control processes that had previously been distributed across three computers were consolidated onto a single testbed server.

A closed-loop experiment is initiated from \texttt{thd-controls}, which selects the optical configuration, initializes the required services, and coordinates image acquisition and DM commands. Experiment routines can then generate probes, estimate the focal-plane electric field, compute EFC updates, and produce diagnostic outputs. THD2-specific parameters are supplied by \texttt{thd-controls}, while commands and images are exchanged with the hardware through CATKit2 services.

This separation limits the coupling between hardware interfaces and experimental algorithms. Device changes can be incorporated at the service level, while WFS\&C routines and experiment scripts remain largely unchanged. The upgraded environment also facilitates rapid transitions between projects. Experiments using different coronagraphs or wavefront-control methods can share the same acquisition, calibration, and diagnostic infrastructure, while project-specific code remains confined to the relevant higher-level procedures. This capability was important for the Roman-like and liquid-crystal coronagraph experiments described in Secs.~\ref{sec:roman-coronagraph-studies} and \ref{sec:suppress-vortex-coronagraphs}, which required substantially different optical configurations but could be operated through the same underlying control system.

\subsection{Validation with the Baseline Four-Quadrant Phase-Mask Coronagraph}
\label{subsec:fqpm-validation}

Following the migration, the first objective was to verify that the new infrastructure preserved the established high-contrast performance of THD2. The validation experiment used a FQPM coronagraph, which has served as a well-characterized baseline configuration throughout the history of the testbed. This choice allowed the control system to be evaluated using an optical configuration for which the alignment behavior, calibration procedures, and expected contrast regime were already known. The aim was therefore not to establish a new performance record, but to demonstrate that extensive changes to the computers, hardware interfaces, and control software had not compromised the operation of the facility.

\begin{figure}
    \centering
   \includegraphics[width = \textwidth]{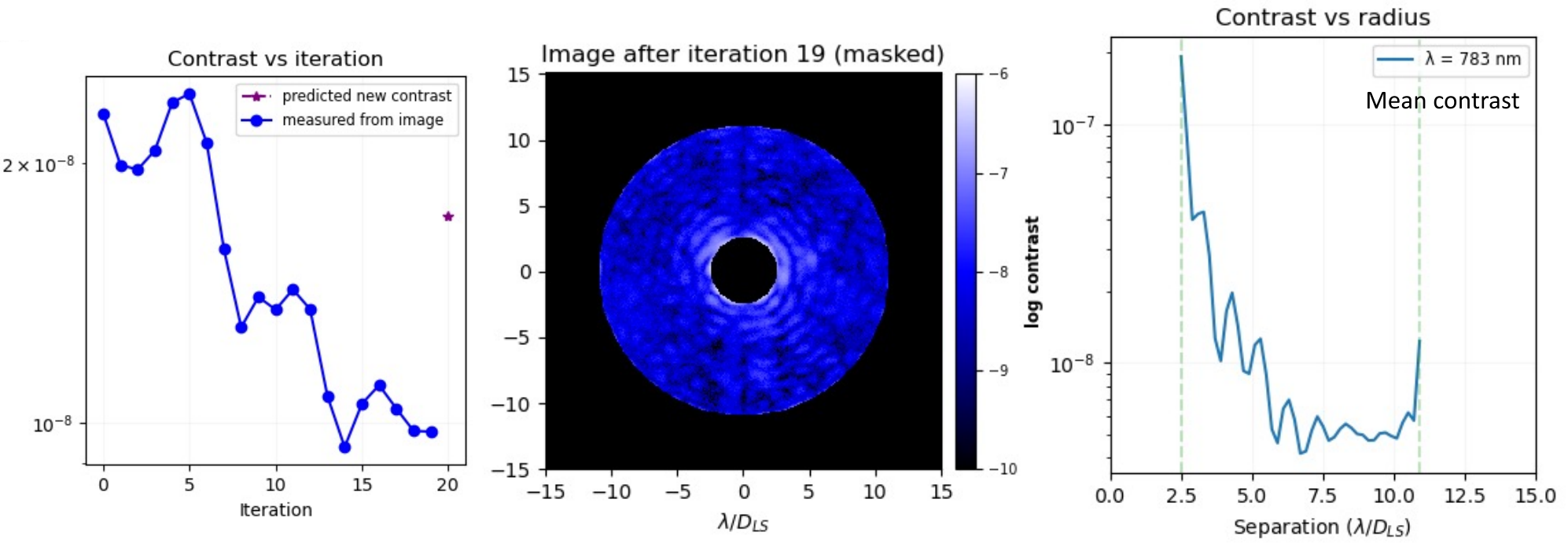}
   \caption[Figure] 
   {\label{fig:fqpm_validation} 
   Validation of the upgraded THD2 infrastructure using the baseline FQPM configuration at $\lambda=783$~nm. Left: evolution of the mean normalized intensity during the EFC loop, showing the contrast measured from each acquired image and the predicted value for the subsequent iteration. This plot captures a closed loop refinement after the initial dark hole has already been established. Center: final coronagraphic image, displayed as normalized log contrast. The annular DH extends from $3$ to $11\,\lambda/D$, and the central region is masked. Right: azimuthally averaged radial contrast profile of the final image, with the dashed lines indicating the DH boundaries. The mean normalized intensity is approximately $9.8\times10^{-9}$ within the controlled region and $5\times10^{-9}$ between 5 and 10 $\lambda$/D.}
   \end{figure}

The experiment was performed in October 2025 in monochromatic light at a central wavelength $\lambda=783$~nm, with the result shown in Fig.~\ref{fig:fqpm_validation}. The focal-plane electric field was estimated through pair-wise probing, and EFC was used to suppress the stellar intensity within an annular DH. The CATKit2-based experiment produced a complete set of diagnostic outputs, including the estimated coherent field, used DM commands, loop convergence, final coronagraphic image, radial contrast profile, and estimate of the incoherent background.

After 19 iterations, the mean normalized intensity within the defined DH region was approximately $9.8\times10^{-9}$. The radial profile reaches $5\times10^{-9}$ between 5 and 10 $\lambda$/D. Closer to the optical axis, we can see a degradation, mostly induced by internal turbulence of the bench. The result recovers the contrast regime previously demonstrated with the FQPM configuration, whose detailed historical performance is reported elsewhere\cite{Potier2020_THD2}. The significance of this measurement is therefore the continuity of performance across the infrastructure migration: THD2 retained its established optical and wavefront-control capability while gaining a maintainable, open-source-based, version controlled and more flexible operating environment.

This validation provided the basis for proceeding to less established configurations. In particular, it demonstrated that coronagraph changes, new control strategies, and broadband experiments could be evaluated within a common framework and accompanied by consistent diagnostic data. The following sections present two examples of the scientific work enabled by this infrastructure.

\section{Enabling Science Case I: Roman Coronagraph Instrument Studies}
\label{sec:roman-coronagraph-studies}

\subsection{The Roman Coronagraph and Community Participation Program}
\label{subsec:roman-cpp}

The Roman Coronagraph Instrument (CGI) will demonstrate coronagraphic WFS\&C with DMs in space and has a principal technical demonstration objective of reaching a contrast of order $10^{-7}$ within a specified DH and observing band\cite{Mennesson2022TheRomanSpace}. Beyond this formal requirement, CGI will provide the first opportunity to exercise several technologies and operational concepts that are directly relevant to future space-based HCI missions.

A central operational constraint is that the CGI WFS\&C loop will be performed with the ground in the loop\cite{Fluckiger2025RomanSpace,Cady2025HighOrderWavefront}. Probe images acquired by the spacecraft are transmitted to the ground, where the focal-plane electric field is estimated and new DM commands are calculated before being sent back to the observatory. The latency and observing time associated with this sequence make the efficiency and robustness of the electric-field estimation strategy operationally important. Improvements that reduce the number of required images or improve the accuracy of the estimate can therefore translate into more efficient use of the available observing time.

The Roman Coronagraph Community Participation Program (CPP) brings together an international community to prepare and investigate observing and data-analysis strategies for CGI. Within the CPP, the Hardware Working Group focuses on instrument operating concepts, including approaches that could improve the efficiency or scientific return of the coronagraphic observations. The THD2 contribution has concentrated on alternative diversity probes for pair-wise focal-plane electric-field estimation. A variety of DH experiments has been run on the testbed to investigate the estimation efficiency and closed-loop behavior of various DM probes. This work has directly informed planned in-orbit observations on Roman, and there is intent to continue using THD2 for laboratory validations of enhanced concepts for Roman CGI as the mission carries on. The work described below builds on the Roman-like HLC experiments published by Laginja et al. (2025)\cite{Laginja2025ExtendedLinearity}, while adding more recent developments carried out in preparation for Roman observations\cite{delaye_romanprobes2026}.

\subsection{Roman-Like Coronagraph Implementations on THD2}
\label{subsec:roman-like-coronagraphs}

A Roman-like HLC configuration was previously developed for THD2 to provide an experimentally representative environment in which CGI WFS\&C strategies could be compared\cite{Laginja2025ExtendedLinearity}. After decommissioning of the dedicated HCI testbed for Roman at NASA/JPL, THD2 remains the only facility with a currently operable Roman-like HLC. The configuration includes a pupil mask, Lyot stop, and partially transmissive focal-plane mask adapted to the THD2 optical geometry and DM actuator count. The focal-plane mask is partially transmissive over a diameter of approximately $5.6\,\lambda/D$, with a phase and amplitude response chosen to reproduce the relevant behavior of the CGI HLC at the central wavelength. The system is not intended to reproduce the complete Roman optical train, but it provides a controlled hardware platform with a comparable coronagraphic response.

\begin{figure}
    \centering
   \includegraphics[width = \textwidth]{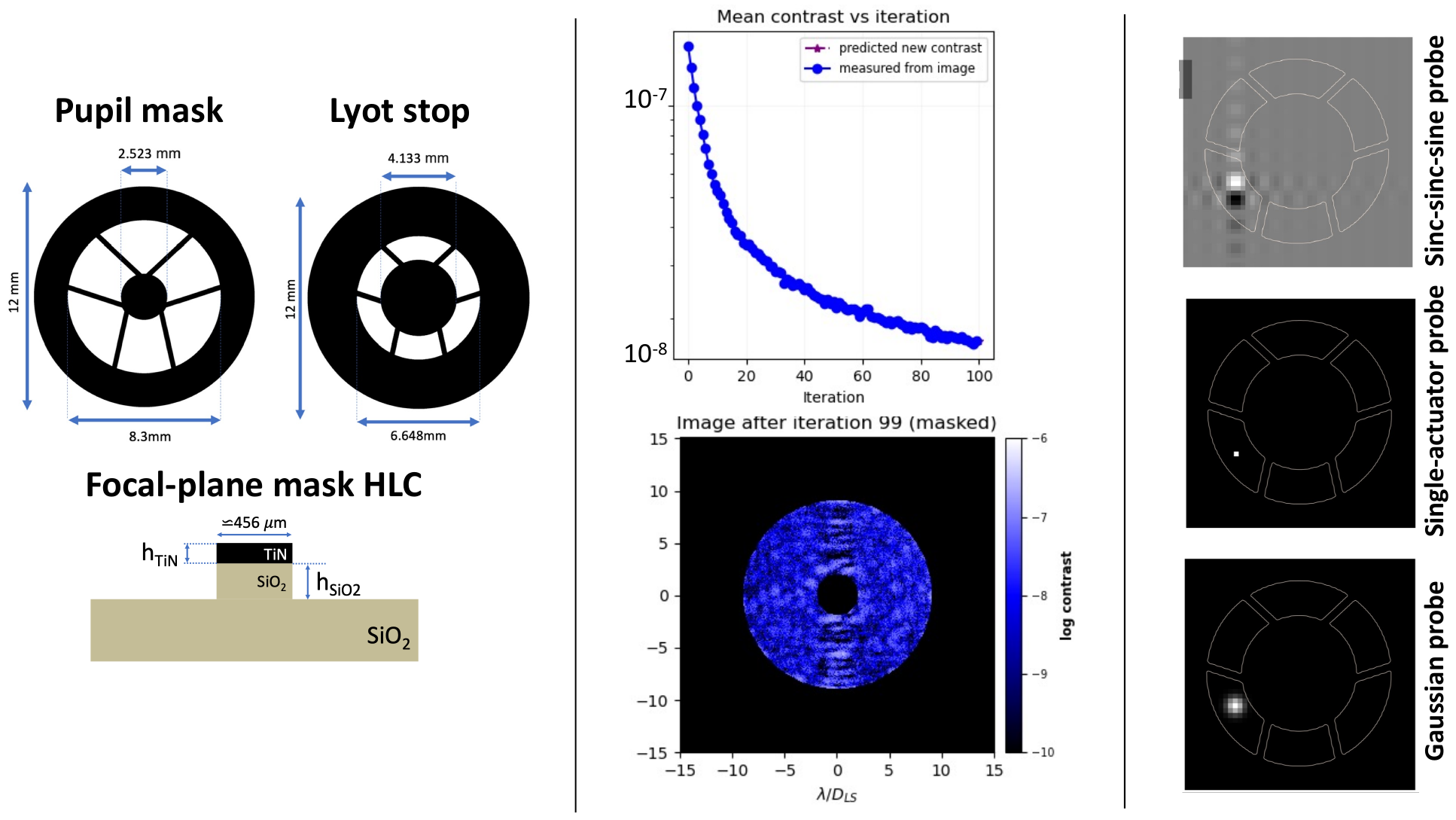}
   \caption[Figure] 
   {\label{fig:roman_hlc_overview} 
   Overview of the Roman-like HLC configuration and pair-wise probe families implemented on THD2. Left: pupil mask, Lyot stop, and cross-section of the partially transmissive HLC focal-plane mask, adapted to the THD2 optical geometry. (Image credit: Laginja et al. 2025\cite{Laginja2025ExtendedLinearity}) Center: representative experimental DH after 99 EFC iterations, shown as normalized log contrast, from $3-10~\lambda/D$, at a wavelength of 780~nm, reaching a contrast of $2\times10^{-8}$ in 99 iterations. Right: representative focal-plane diversity patterns produced by the THD2 for the Roman baseline sinc--sinc--sine probe (top), a single-actuator probe (middle), and a Gaussian probe (bottom). The overlaid contour indicates the Lyot stop of Roman CGI in this coronagraph configuration.}
   \end{figure}

Closed-loop experiments with this configuration reached a mean normalized intensity of approximately $2\times10^{-8}$ in the controlled region. The detailed mask design, numerical modeling, calibration procedure, and experimental performance are presented in Laginja et al. (2025)\cite{Laginja2025ExtendedLinearity}. Here, this configuration is used primarily to demonstrate that the upgraded THD2 infrastructure can support Roman-relevant WFS\&C experiments. Figure~\ref{fig:roman_hlc_overview} summarizes the HLC configuration, a representative experimental DH with it convergence plit, and the progression of probe concepts discussed in the following sections.

\subsection{Enhanced Pair-Wise Probing Strategies}
\label{subsec:enhanced-pair-wise-probing}

Pair-wise probing estimates the coherent focal-plane electric field by applying pairs of opposite DM commands and measuring the corresponding difference in intensity\cite{Giveon2011PairwiseDeformableMirror}. The imposed phase diversity must produce a sufficiently strong response throughout the DH while remaining within the validity range of the estimator. Probe morphology and amplitude therefore influence both the signal-to-noise ratio of the measurement and the magnitude of neglected non-linear terms\cite{Groff2016MethodsLimitationsFocal}.

\subsubsection{Baseline Sinc--Sinc--Sine Probes and Single-Actuator Probes}
\label{subsec:baseline-and-simplified-probes}

The CGI baseline probe family is constructed from combinations of sinc functions and sinusoidal modulation chosen to illuminate the spatial frequencies corresponding to the DH\cite{Cady2025HighOrderWavefront}. The functional form of these probes was retained in the THD2 experiments, with the commands adapted from the Roman DM geometry to the $32\times32$ actuator DM used on THD2.

Although these extended probe shapes provide broad modulation of the DH, previous THD2 work showed that comparably useful electric-field diversity can be generated by commanding a single DM actuator\cite{Potier2020_THD2}. Laginja et al. (2025)\cite{Laginja2025ExtendedLinearity} compared the baseline sinc--sinc--sine probes with single-actuator and other simplified probe families in simulation and on the THD2 HLC configuration. Their localized DM command also reduced the influence of unmodeled behavior distributed across many actuators.

\subsubsection{Gaussian Probes and Nonlinearity Considerations}
\label{subsec:gaussian-probes}

Following the published THD2 study, the CPP Hardware Working Group investigated smooth Gaussian DM commands as an additional probe family\cite{delaye_romanprobes2026}. The Gaussian probes are placed in approximately the same region of the DM as the baseline sinc probes and produce a similar focal-plane diversity, but they command fewer actuators with large displacements. Their morphology is also similar to a sinc--sinc probe without the additional sinusoidal modulation, which was the best-performing member of the baseline probe set in the non-linearity analysis (see also Fig.~2 in Delaye et al. 2026\cite{delaye_romanprobes2026}).

The standard pair-wise estimator is derived using a linear approximation of the differential intensity. For sufficiently small probe amplitudes, the neglected higher-order contributions remain small. As the probe amplitude increases, however, these terms become significant and their magnitude depends on the spatial structure of the DM probe command. This contribution can be quantified by comparing the fraction of the linear term versus the non-linear terms in the PW probing expansion of the measured modulation (Eq.~13 in Delaye et al. 2026\cite{delaye_romanprobes2026}). At equal focal-plane probe intensity, the Gaussian probes produce a lower fraction of non-linear error than two of the three baseline sinc--sinc--sine probes over the investigated range.
This comparison suggests that the additional phase modulation by a sine/cosine on the sinc probes introduces higher-order terms without providing a corresponding increase in useful estimated signal.

\subsubsection{Implications for Coherent Differential Imaging and On-Sky Use}
\label{subsec:cdi-and-on-sky-use}

The accuracy of pair-wise electric-field estimates is relevant not only during DH generation, but also for coherent differential imaging\cite{Potier2026CoherentDifferential,delaye_romanprobes2026} (CDI). CDI uses the estimated coherent field to distinguish residual stellar speckles from incoherent signals in the science image. Biases introduced by probe non-linearities can therefore propagate into both the WFS\&C loop and the subsequent data analysis. Probe shapes with smaller higher-order contributions are expected to provide more reliable coherent estimates, particularly when strong probes are required to obtain adequate signal-to-noise ratio.

The Gaussian-probe results motivated the inclusion of alternative probe shapes in a planned Roman observing campaign after CGI commissioning, where baseline and alternative probes will be compared in flight in terms of estimation accuracy and operational efficiency. This activity, developed within the CPP, extends the progression from simplified probes on THD2 to Roman-like hardware experiments and ultimately to an on-orbit observing concept. In this way, THD2 serves as a bridge between component testing, algorithm development and mission operations: its Roman-like HLC provides a representative optical configuration, while the upgraded CATKit2 infrastructure enables probe families and control sequences to be exchanged and evaluated within a common experimental framework. As the Roman mission launches and carries on, the THD2 team plans to support the validation of further CGI enhancements prepared by the CPP.

\section{Enabling Science Case II: SUPPPPRESS and Liquid-Crystal Vortex Coronagraphs}
\label{sec:suppress-vortex-coronagraphs}

    \begin{figure}
    \centering
   \includegraphics[width = \textwidth]{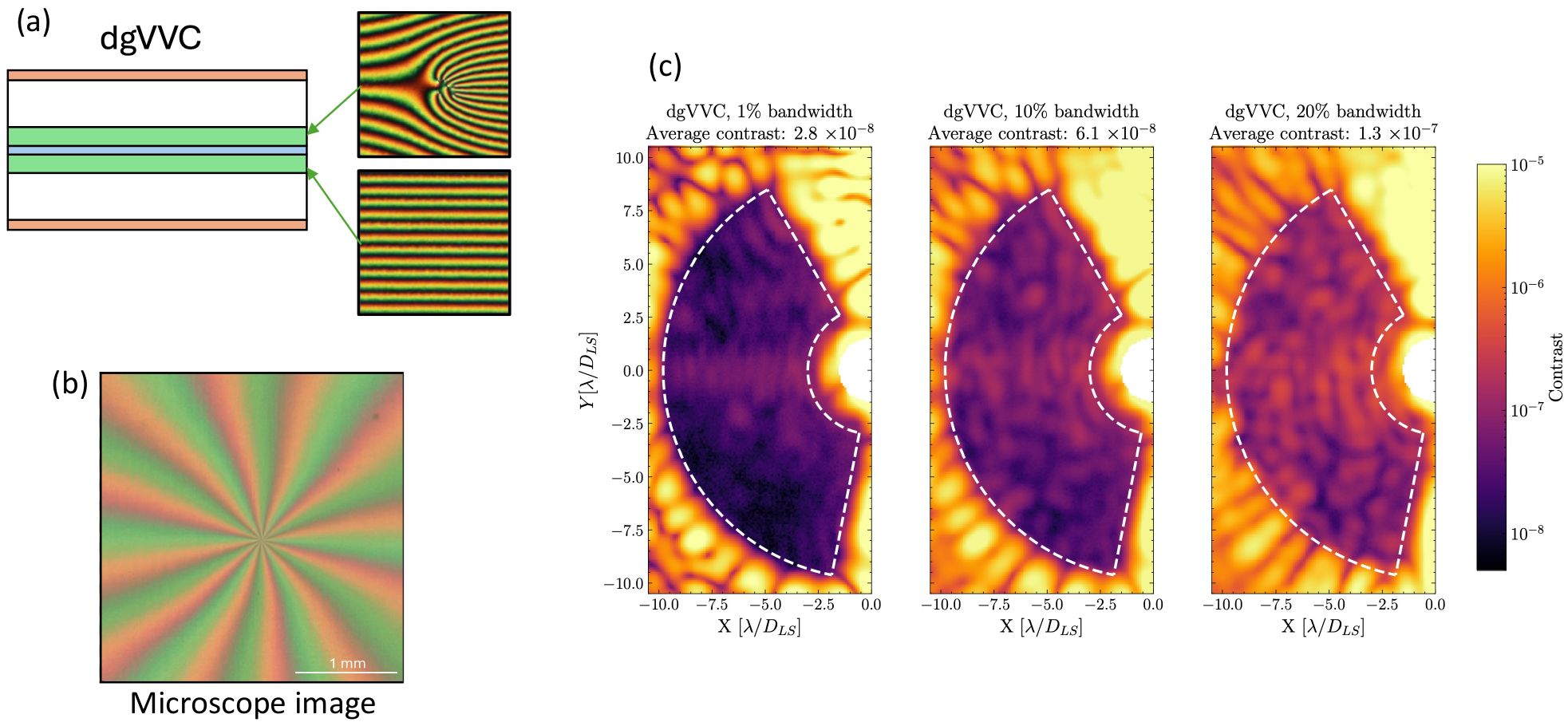}
   \caption[Figure] 
   {\label{fig:suppppress_figure} 
   Design and experimental performance of the charge-6 double-grating vector vortex coronagraph (dgVVC) developed within SUPPPPRESS. Credit: Landman \& Doelman et al. (2026)\cite{LandmanDoelman2026SUPPPPRESS}. (a) Schematic of the dgVVC architecture, combining a forked charge-6 grating with a polarization grating; the insets show the corresponding liquid-crystal patterns. (b) Microscope image of the fabricated dgVVC pattern, with a 1~mm scale bar. (c) Experimental DHs obtained with th edgVVC on the upgraded THD2 testbed in 1\%, 10\%, and 20\% fractional bandwidths centered near $\lambda=780$~nm. The dashed contours indicate the one-sided scoring region used to exclude ghost-contaminated areas. The mean normalized intensities within this region are $2.8\times10^{-8}$, $6.1\times10^{-8}$, and $1.3\times10^{-7}$, respectively. All three DH images are displayed using the same logarithmic contrast scale.}
   \end{figure}

\subsection{The SUPPPPRESS Project}
\label{subsec:suppress-project}

SUPPPPRESS was a two-year ESA-funded activity led by Leiden University and NOVA, with contributions from SRON, cosine and LIRA/Paris Observatory-PSL\cite{LandmanDoelman2026SUPPPPRESS,Landman2026suppppress}. The project investigated liquid-crystal phase-mask technology for future space-based HCI instruments, including the manufacturing and characterization of individual components, assembly of multi-grating VVCs and simple VVCs, environmental testing, and system-level HCI demonstrations. THD2 provided the testbed on which the completed components were evaluated with active focal-plane WFS\&C. The detailed mask design, liquid-crystal patterning, manufacturing, metrology, and environmental qualification are presented by Landman \& Doelman et al. (2026)\cite{LandmanDoelman2026SUPPPPRESS}. In the present work, we recall the experimental results obtained on THD2.

\subsection{Experimental Setup and Results}
\label{subsec:suppress-experiments-results}

The experiments used the tunable broadband source installed on THD2 within the SUPPPPRESS project and were centered near $\lambda=780$~nm. The focal-plane electric field was estimated with pair-wise probing, and EFC was used to generate a DH over approximately $3$--$10\,\lambda/D$. Measurements were obtained in several fractional bandwidths to evaluate the combined chromatic behavior of the coronagraph and the WFS\&C system.

The experimental configuration included linear polarizers and quarter-wave plates to select a single circular-polarization state. These optics introduced identifiable ghost reflections in the focal plane. The ghosts remained at fixed detector locations when the coronagraph or observing wavelength was changed, confirming that they are distinct from residual coronagraphic leakage. To prevent these artifacts from biasing the reported performance, the quantitative analysis was restricted to a one-sided DH region that excluded the affected area.

Figure~\ref{fig:suppppress_figure} summarizes the dgVVC, manufactured in SUPPPPRESS. The mean normalized intensity within the selected DH reached $2$--$3\times10^{-8}$ in a 1\% bandwidth. The measured value increased to approximately $6\times10^{-8}$ in a 10\% bandwidth and to approximately $10^{-7}$ in a 20\% bandwidth. The conventional VVC (see Fig.~13 in Landman \& Doelman et al. (2026)\cite{LandmanDoelman2026SUPPPPRESS}) showed a similar broadband trend, with mean normalized intensities of order $5\times10^{-8}$, $7\times10^{-8}$, and $10^{-7}$ in 5\%, 10\%, and 20\% bandwidths, respectively. Exact values, scoring regions, and the full component comparison are reported by Landman \& Doelman et al. (2026)\cite{LandmanDoelman2026SUPPPPRESS}.

\section{Toward a European HCI Facility}
\label{sec:european-hci-facility}

\subsection{THD2 as a Shared European Infrastructure}
\label{subsec:shared-european-infrastructure}

The upgraded control infrastructure enables THD2 to be operated increasingly as a shared facility rather than as a sequence of independently developed experiments. Common hardware interfaces, calibration procedures, and diagnostic products allow several users to understand and operate the testbed without reproducing its low-level control environment. Experiments can consequently be prepared and executed back-to-back, while developments from one project can be retained for later users. Experienced operators remain essential for alignment, calibration, and interpretation, but the number of procedures that depend exclusively on individual expertise is reduced.

THD2 provides a complementary European in-air platform within the wider international network of HCI testbeds. Its established partners include CNES, NOVA, LAM, Lagrange/OCA, and STScI, while the Roman and SUPPPPRESS activities presented here demonstrate its ability to support collaborations beyond its host institute. Recently accepted European\footnote{ERC ECHOES project \url{https://lira.obspm.fr/erc-echoes/}} and French\footnote{France 2030 AMINO project: \href{https://pepr-origins.fr/en/projet/advanced-multi-spectral-imaging-using-novel-wavefront-sensors-with-optical-photon-counting-detectors-amino-2/}{website}.} funding will contribute to further developments of the facility, including more sensitive and efficient WFS\&C, polychromatic control, and investigations of post-processing limitations in a controlled environment. A principal objective of this next phase is to coordinate access to THD2 and connect European technology developments with the broader international preparation for HWO.

\subsection{CATKit2-HCI: From Hardware Control to Algorithm Sharing}
\label{subsec:catkit2-hci}

CATKit2 provides the generic services required to control and monitor testbed hardware, but many higher-level HCI procedures are also common between facilities. The CATKit2-HCI collaboration extends the shared-software approach to algorithms and experimental workflows, including pair-wise probing, EFC, DH definitions, regularization strategies, and standardized diagnostic products\cite{Soummer2026}. Established methods can therefore be maintained as reusable implementations rather than being independently rewritten for each experiment or testbed.

This common layer supports reproducibility and more direct comparisons between facilities. When the same algorithmic implementation is applied to different systems, observed differences can be attributed more confidently to the hardware, calibration, or environment rather than to unrelated software choices. Testbed-specific models and conventions remain necessary, but can be incorporated through defined interfaces while the general WFS\&C workflow is shared. THD2 is contributing to this effort and provides a European environment in which these common tools can be validated across multiple coronagraphs and science cases.

\subsection{Community Building and Workshops}
\label{subsec:community-building}

The facility and software developments described above form part of a broader effort to coordinate European research and development for space-based HCI. A workshop series initiated in 2024 has brought together researchers working across exoplanet science, coronagraphy, WFS\&C, DMs, detectors, telescope optics, post-processing, nulling interferometry, and system engineering, while also considering relevant synergies with ground-based HCI. The three meetings were held in Paris in 2024\cite{Laginja2025AdvancingEuropeanHCI},\footnote{\url{https://hcieurope.sciencesconf.org/}} Heidelberg in 2025\cite{Chauvin2026Continuing},\footnote{\url{https://hcieurope-mpia.sciencesconf.org/}} and Edinburgh in 2026.\footnote{\url{https://www.roe.ac.uk/workshop/sb-hci-euro-3/}}

The series progressively moved from identifying European expertise and scientific and technological priorities toward defining coordinated development activities. A principal outcome of the latest meeting was the formation of a European consortium to investigate a potential near-infrared coronagraph contribution to HWO. The discussions also identified the value of European near-infrared system-level demonstration capabilities for maturing relevant technologies, of which THD2 and its future developments may form one element.

\section{Conclusions}
\label{sec:conclusions}

We have presented the modernization of the THD2 HCI testbed through its migration to the open-source CATKit2 control framework\cite{Catkit2_2026}. The upgrade replaced commercial, outdated software, consolidated processes previously distributed across three computers, and introduced a common, version-controlled interface for hardware operation and experiment development. Validation with the baseline FQPM configuration recovered a mean normalized intensity of $5\times10^{-9}$ between 5 and 10 $\lambda$/D, demonstrating that the established optical performance of THD2 was preserved across the infrastructure migration.

The scientific value of the upgrade was illustrated through two subsequent projects. For Roman CGI studies, the upgraded environment supported Roman-like HLC experiments and comparisons of alternative pair-wise probe families\cite{Laginja2025ExtendedLinearity}. The more recent Gaussian probes produced lower fractional non-linear errors than the sinusoidally modulated baseline probes under comparable conditions, motivating their inclusion in a planned alternate-probing observation after CGI commissioning. Within the ESA-funded SUPPPPRESS project, newly manufactured liquid-crystal VVCs were integrated and evaluated with active broadband WFS\&C\cite{LandmanDoelman2026SUPPPPRESS}. The charge-6 dgVVC reached mean normalized intensities of approximately $2$--$3\times10^{-8}$ in a 1\% bandwidth and of order $10^{-7}$ in a 20\% bandwidth. These measurements also identified the combined effects of polarization leakage, optical ghosts, coronagraph manufacturing errors, and chromatic WFS\&C as priorities for further development.

Together, these results demonstrate that the principal benefit of the upgrade extends beyond replacing obsolete infrastructure. CATKit2 enables different coronagraphs, control strategies, and collaborations to use a common experimental environment, improving reproducibility and reducing the effort required to transition between projects. Further development through CATKit2-HCI\cite{Soummer2026}, coordinated access to THD2, and broader community activities for space-based HCI will support the continued use of the testbed as a platform for maturing technologies and observing concepts relevant to Roman, HWO, and future direct-imaging missions.

\acknowledgments 
This work benefited from the CATKit2-HCI collaboration which originated at the HiCAT testbed at STScI \cite{Soummer2026}.
This work benefited from the Action Sp\'ecifique Haute R\'esolution Angulaire(ASHRA) of CNRS/INSU, co-funded by CNES.
The development of the THD2 test bench was partly supported by Centre National d'\'Etudes Spatiales (CNES) through R\&D fundings R-S14/SU-002-068, R-S17/SU-0002-068, R-S19/SU-0002-105.
This work was also supported by SUPPPPRESS, funded by the European Space Agency (ESA) under the tender number \linebreak TDE-TEC-MMO AO/1-11613/23/NL/AR.
Iva Laginja acknowledges partial support from a postdoctoral fellowship issued by the Centre National d’Etudes Spatiales (CNES) in France.
Emiel H. Por was supported in part by the NASA Hubble Fellowship grant HST-HF2-51467.001-A awarded by the Space Telescope Science Institute, which is operated by the Association of Universities for Research in Astronomy, Incorporated, under NASA contract NAS5-26555.
The authors acknowledge the use of AI tools (GPT-5.6 Sol) for improvements and cleanup of language and grammar.

\bibliography{references}
\bibliographystyle{spiebib}

\end{document}